# Calculation of excitation function of some structural fusion material for (n,p) reactions up to 25 MeV

Tarik S.RESHID[1]
[1]Faculty of Basic Education, Department of General Science, Salahaddin University, Erbil, Iraq
E-mail: (taqa@hotmail.com)

Abstract

Fusion serves an inexhaustible energy for humankind. Although there have been significant research and development studies on the inertial and magnetic fusion reactor technology, Furthermore, there are not radioactive nuclear waste problems in the fusion reactors. In this study, (n, p) reactions for some structural fusion materials such as $^{27}$Al, $^{51}$V, $^{52}$Cr, $^{55}$Mn and $^{56}$Fe have been investigated. The new calculations on the excitation functions of $^{27}Al(n,p)^{27}Mg$, $^{51}V(n,p)^{51}Ti$, $^{52}Cr(n,p)^{52}V$, $^{55}Mn(n,p)^{55}Cr$ and $^{56}Fe(n,p)^{56}Mn$ reactions have been carried out up to 30 MeV incident neutron energy. Statistical model calculations, based on the Hauser-Feshbach formalism, have been carried out using the TALYS-1.0 and were compared with available experimental data in the literature and with ENDF/B-VII, T=300k; JENDL-3.3, T=300k and JEFF3.1, T=300k evaluated libraries .

## 1. Introduction:

The experimental cross-sections can be extensively used for the investigation of the structural materials of the fusion reactors, radiation damage of metals and alloys, tritium breeding ratio, neutron multiplication and nuclear heating in the components, neutron spectrum, and reaction rate in the blanket and neutron dosimetry (1–4). And also these obtained data are necessary to develop more nuclear theoretical calculation models in order to explain nuclear reaction mechanisms (5–8). These nuclear reaction models are frequently needed to provide the estimation of the particle–induced reaction cross–sections, especially if the experimental data are not obtained or on which they are hopeless to measure the cross–sections; due to the experimental difficulty. So the cross-section evaluation for materials irradiated by neutrons attaches special importance to use of systematics of neutron induced reaction cross-section. Such predictions can guide the design of the target/blanket configurations and can reduce engineering over design costs.

The nuclear physics community has developed tools for specific applications, such as acceleratordriven systems, which can shed light on the many approximations in nuclear applications. One of these tools is the modern reaction code called TALYS (9-12). TALYS is a software for the simulation of nuclear reactions, which includes many state-of the- art nuclear models to cover all main reaction mechanisms encountered in light particle-induced nuclear reactions.TALYS provides a complete description of all reaction channels and observables and in particular takes into account all types of direct, pre-equilibrium, and compound mechanisms to estimate the total reaction probability as well as the competition between the various open channels. The code is optimized for incident projectile energies, ranging from 1 keV up to 200 MeV on target nuclei with mass numbers between 12 and 339. It includes photon, neutron, proton, deuteron, triton, 3He, and particles as both projectiles and ejectiles, and single-particle as well as multi-particle emissions and fission. All experimental information on nuclear masses, deformation, and low-lying states spectra is considered, whenever available. If not, various local and global input models have been incorporated to represent the nuclear structure properties, optical potentials, level densities, -ray strengths, and fission properties. The TALYS code was designed to calculate total and partial cross sections, residual and isomer production cross sections, discrete and continuum -ray production cross sections, energy spectra, angular distributions,double-differential spectra, as well as recoil cross sections.

Neutron irradiation produces significant changes in the mechanical and physical properties of each of structural fusion material systems raising feasibility questions and design limitations. A focus of the research and development effort is to understand these effects, and through the development of specific compositions and microstructures, produce materials with improved and adequate performance (9, 10). $^{27}$Al, $^{51}$V, $^{52}$Cr, $^{55}$Mn and $^{56}$Fe nuclei are the some structural fusion materials (11, 12). Nuclear data evaluation is generally carried out on the basis of experimental data and theoretical model calculations.

It is both physically and economically impossible to measure necessary cross-section for all the isotopes in the periodic table for a wide range of energies. Therefore model calculations play an important role in the evaluation of nuclear data (13-15) . In the present paper, by using equilibrium and pre-equilibrium reaction mechanisms, the (n,p) reactions for some structural fusion materials were investigated up to 25 MeV.

## 2. Nuclear reaction model

Most nuclear calculations adopt nuclear reaction rates evaluated by the HF model. It relies on the fundamental assumption (Bohr hypothesis) that the capture process occurs by means of the intermediary production of a compound system that can reach a state of thermodynamic equilibrium. This compound system is then classically referred to as the compound nucleus (CN). The formation
Of a CN occurs if the CN level density, at the excitation energy corresponding to the projectile incident energy, is sufficiently high. The reaction $I^{\mu} + a \rightarrow I' + a'$ represents the capture of a light particle a onto a target I in its state μ, where μ = 0 corresponds to the target ground state, leaving the residual nucleus I′ and the particle or photon a′; TALYS estimates the corresponding cross section by the compound nucleus formula for the binary cross section in the full jls scheme, i.e.

$$\sigma^{\mu}_{\alpha\alpha'} = D^{comp}\pi\lambda^2 \sum_{J=mod(I^{\mu}+s,1)}^{l_{max}+I^{\mu}+s} \sum_{II=-1}^{1} \frac{2J+1}{(2I^{\mu}+1)(2s+1)}$$

$$\sum_{j=|J-I^{\mu}|}^{J+I^{\mu}} \sum_{l=|j-s|}^{j+s} \sum_{j'=|J-I'|}^{J+I'} \sum_{l'=|j'-s'|}^{J'+s'} \delta_{\pi}(\alpha)\delta_{\pi}(\alpha')$$

$$\frac{\langle T^{J}_{\alpha lj}(E_{a'})\rangle\langle T^{J}_{\alpha' l' j'}(E_{a'})\rangle}{\sum_{\alpha'',l'',j''}\delta_{\pi}(\alpha'')\langle T^{J}_{\alpha'',l'',j''}(E_{a''})\rangle} W^{J}_{\alpha lj\alpha' l' j'} \quad .....(1)$$

In the above equations, Ea, s, $\pi_0$, $l$, and j represent the projectile energy, spin, parity, orbital, and total angular momentum, respectively, and $l_{max}$ is the maximum projectile $l$-value. The same symbols but labelled by a prime correspond to the ejectile. The symbols $I^{\mu}$ , $II^{\mu}$ (I',II')represent the spin and parity of the target (residual) nucleus, while J and II correspond to the spin and parity of the compound system. The initial system of projectile and target nucleus is designated by α = {a, s, Ea, $E^{\mu}_{x}$ , $I^{\mu}$, $II^{\mu}$} where $E^{\mu}_{x}$ corresponds to the excitation energy of the target nucleus. $\alpha'$ = {a', s', $E_{a'}$ ,$E'_{x}$ , $I'$,$II'$} is similar for the final system of ejectile and residual nucleus. $\delta_{\pi}(\alpha)$ = 1, if $(-1)^{l}\,\pi_0 II^{\mu} = II$ and 0 otherwise (the same holds for the final system $\alpha'$). In Eq. 1, $\lambda$ is the relative motion wave length, T the transmission coefficient, W the width fluctuation correction factor and $D^{comp}$ the depletion factor given by

$$D^{comp} = \frac{\left[\sigma_{reac} - \sigma^{dic,direct} - \sigma^{PE}\right]}{\sigma_{reac}} \ldots\ldots\ldots (2)$$

where $\sigma_{rec}$ is the total reaction cross section, $\sigma^{disc,direct}$ is the total discrete direct cross section, and $\sigma^{PE}$ is the preequilibrium cross section. It is assumed by the TALYS code that direct and compound contributions can be added incoherently. The formula for $D^{comp}$ is only applied to weakly coupled channels that deplete the flux, such as contributions from the direct or pre-equilibrium processes. For deformed nuclides, the effect of direct transitions to discrete levels is included directly in the coupled-channels scheme and the $\sigma^{disc,direct}$ term is omitted from Eq. 2.The HF formalism is valid only if the formation and decay of the CN are totally independent. This so-called Bohr hypothesis may not be fully satisfied, particularly in cases where a few strongly and many weakly absorbing channels
are mixed. As an example, the HF equation is known to be invalid when applied to the elastic channel, since in that case the transmission coefficients for the entrance and exit channels are identical, and hence correlated. These correlations enhance the elastic channel and accordingly decrease the other open channels. To account for these deviations, a width fluctuation correction W is introduced into the HF formalism (see Eq. 1). When many competing channels are open, above a few MeV of projectile energy, the width fluctuation correction can be neglected. Each transmission coefficient T is estimated for all levels with experimentally known energy, spin, and parity. In that case, we simply have$\langle T^{J}_{\alpha' l' j'}(E_{a'})\rangle = T^{J}_{\alpha' l' j'}(E_{a'})$. However, if the excitation energy $E'_x$ , which is implicit in the definition of channel $\alpha'$, corresponds to a state in the continuum, we have an effective transmission coefficient for an excitation energy bin of width $\Delta E_x$ determined by the integral

$$\langle T^{J}_{\alpha' l' j'}(E_{a'})\rangle = \int_{E'_x - \frac{1}{2}\Delta E_x}^{E'_x + \frac{1}{2}\Delta E_x} dE_y \rho\big(E_y, J, II\big) T^{J}_{\alpha' l' j'}(E_{a'}) \ldots\ldots\ldots\ldots\ldots\ldots\ldots\ldots\ldots\ldots\ldots\ldots..(3)$$

Over the level density $\rho(E_y, J, II)$ , at an energy $E_y$, spin J,and parity II in the CN or residual nucleus. For increasing energy or nuclei for which the CN does not have time to reach thermodynamic equilibrium, direct or pre-equilibrium processes may become significant. In TALYS, the direct component is derived from the Distorted Wave Born Approximation for spherical nuclei and the coupled-channels equations for deformed nuclei. The preequilibrium emission can occur after the first stage of the reaction but long before statistical equilibrium of the CN is reached. Although pre-equilibrium processes can cover asizable part of the reaction cross section for intermediate energies (typically above a few MeV in stable nuclei). Both classical and quantum mechanical pre-equilibrium models exist and are included in TALYS. One of the most widely used pre-equilibrium models is the (one- or two-component) exciton model(16) , in which the nuclear state is characterized at any moment during the reaction by the total energy$E^{tot}$ and the total number of particles above and holes below the Fermi surface. Particles (p) and holes (h) are referred to as excitons. Furthermore, it is assumed that all possible ways of sharing the excitation energy between different particle-hole configurations at the same exciton number n = p + h have an equal priori probability. To monitor the evolution of the scattering process, one merely traces the temporal development of the exciton number, which changes in time as a result of intranuclear two-body collisions. The basic starting point of the exciton model is a time-dependent master equation, which describes the probability of transitions to more and less complex particle-hole states as well as transitions to the continuum (emission). Upon integration over time, the energy-averaged emission spectrum is derived. These assumptions ensure that the exciton model is amenable to practical calculations. The disadvantage, however, is the introduction of a free parameter, namely the average matrix element of the residual two-body interaction, occurring in the transition rates between two exciton states. A thermodynamic equilibrium holds locally to a good approximation inside stellar interiors. Consequently, the energies of both the targets and projectiles, as well as their relative energies E, obey a Maxwell-Boltzmann distribution corresponding to the temperature T at that location (or a blackbody Planck spectrum for photons) . In a thermodynamic equilibrium situation, the relative populations of the various levels of nucleus $I^{\mu}$ with excitation energies $E^{\mu}_x$ obey a Maxwell-Boltzmann distribution. The effective stellar rate per pair of particles in the entrance channel at a temperature T, taking account of the contributions of various target excited states, is finally expressed in a classical notation as

$$N_A\langle\sigma v\rangle^{*}_{\alpha\alpha'}(T) = \left(\frac{8}{\pi m}\right)^{\frac{1}{2}} \frac{N_A}{(KT)^{\frac{3}{2}}\, G_1(T)}$$

$$\int_0^\infty \sum_\mu \frac{(2I^\mu+1)}{(2I^0+1)} \sigma^\mu_{\alpha\alpha'}(E) E Exp\left(-\frac{E+E_x^\mu}{kT}\right) dE \ldots\ldots\ldots\ldots\ldots.. (4)$$

where k is the Boltzmann constant, m the reduced mass of the $I^0$ + a system, $N_A$ the Avogadro number, and

$$G_1(T) \sum_\mu (2I^\mu + 1)/(2I^0 + 1) \exp(-E_x^\mu/kT) \ldots\ldots\ldots\ldots\ldots..(5)$$

the T -dependent normalized partition function.

The uncertainties involved in any cross section calculation are of two origins:

– the first one is related to the reaction mechanism, i.e. the model of formation and de-excitation of the CN itself. Reaction mechanisms have compound, pre-equilibrium, and direct components. The compound formation is described by the HF theory.

– Another type of uncertainty comes from the evaluation of the nuclear quantities required for calculating the transmission coefficients in Eqs. (1) - (4), i.e. the ground and excited state properties (masses, deformations, matter densities, excited state energies, spins, and parities, . . . ), nuclear level densities, -ray strength, optical model potential, and fission properties. When not available experimentally, this information has to be derived from nuclear models. Ideally, when dealing with nuclear applications, the various nuclear ingredients should be determined from *global*, *universal*, and *microscopic* models. The large number of nuclides involved in the modeling of some nucleosynthesis mechanisms implies that global models should be used. On the other hand, a universal description of all nuclear properties within a unique framework for all nuclei involved in a nuclear network ensures the essential coherence of predictions for all unknown data. Finally, a microscopic description provided by a physically sound theory based on first principles ensures extrapolations away from experimentally known energy or mass regions that are likely to be more reliable than predictions derived from more or less parameterized approaches of various types and levels of sophistication.
This is true even as new generations of such models are starting to be developed to compete with more phenomenological highly-parameterized models for the reproduction of experimental data (23-26) Only a few reaction model codes adopt the largest possible extent of global and coherent microscopic (or at least semi-microscopic) models.

Briefly, the TALYS-1.0 code (equilibrium and pre-equilibrium) is optimized for incident projectile energies ranging from 1 keV up to 200 MeV on target nuclei, with mass numbers between 12 and 339. It includes photon, neutron, proton, deuteron, triton, $^3$He and $\alpha$-particles, as well as projectiles and ejectiles and single-particle and multi-particle emissions and fission. Equilibrium and pre-equilibrium particle emissions during the decay process of a compound nucleus are very important for a better understanding of the nuclear reaction mechanism induced by medium energy particles. The highly excited nuclear system produced by charged particles' first decays by emitting fast nucleons at the pre-equilibrium (PE) stage and, later on, by the emission of low-energy nucleons at the equilibrium (EQ) stage. (17-29).

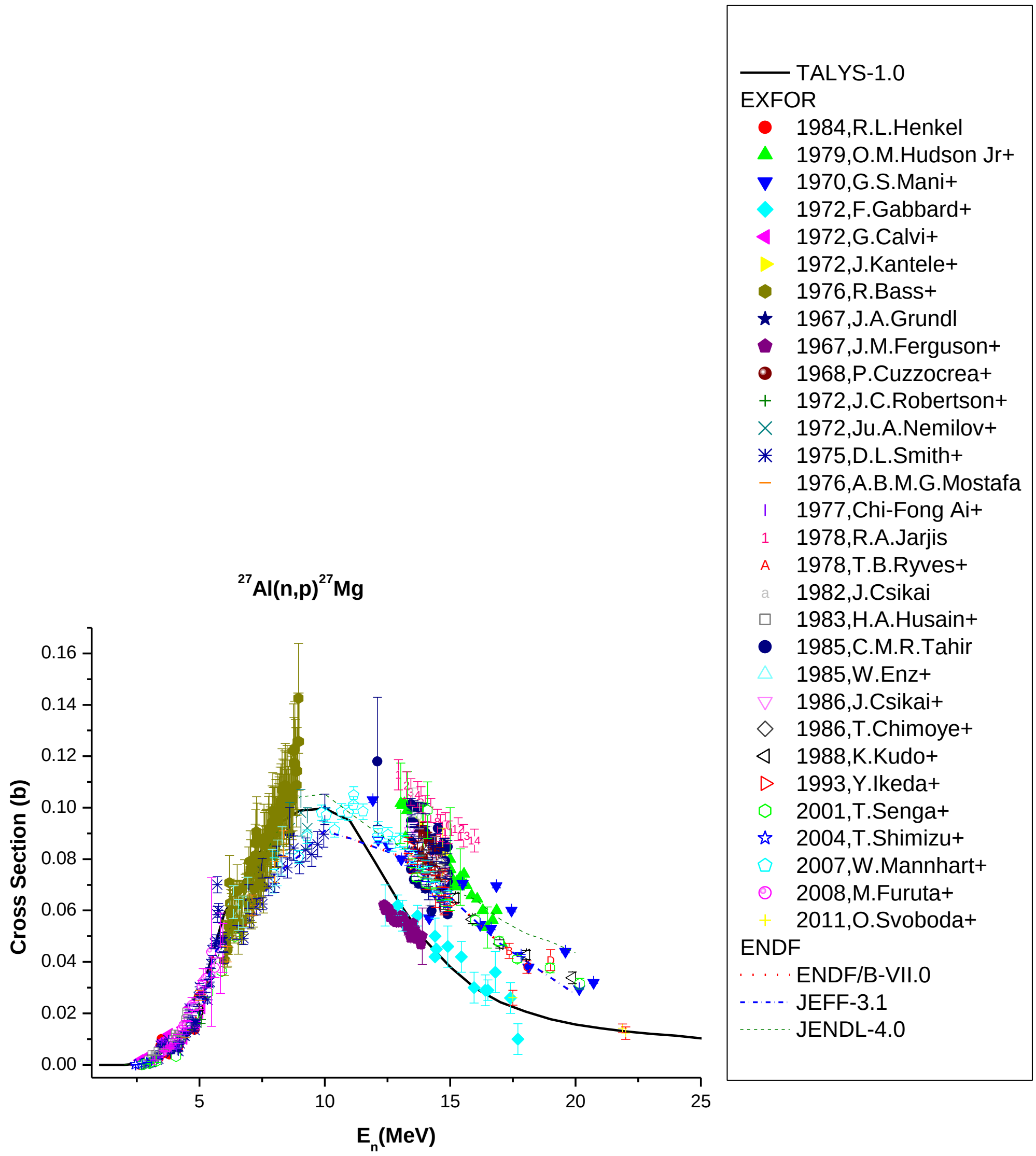

Fig. 1 The comparison of calculated excitation function using TALS-1.2 of $^{27}Al(n,p)^{27}Mg$ reaction with available experimental values and evaluated nuclear data files ENDF/B-VI, JENDL-3.3 and JEFF3.1.The values reported in Ref. (30)

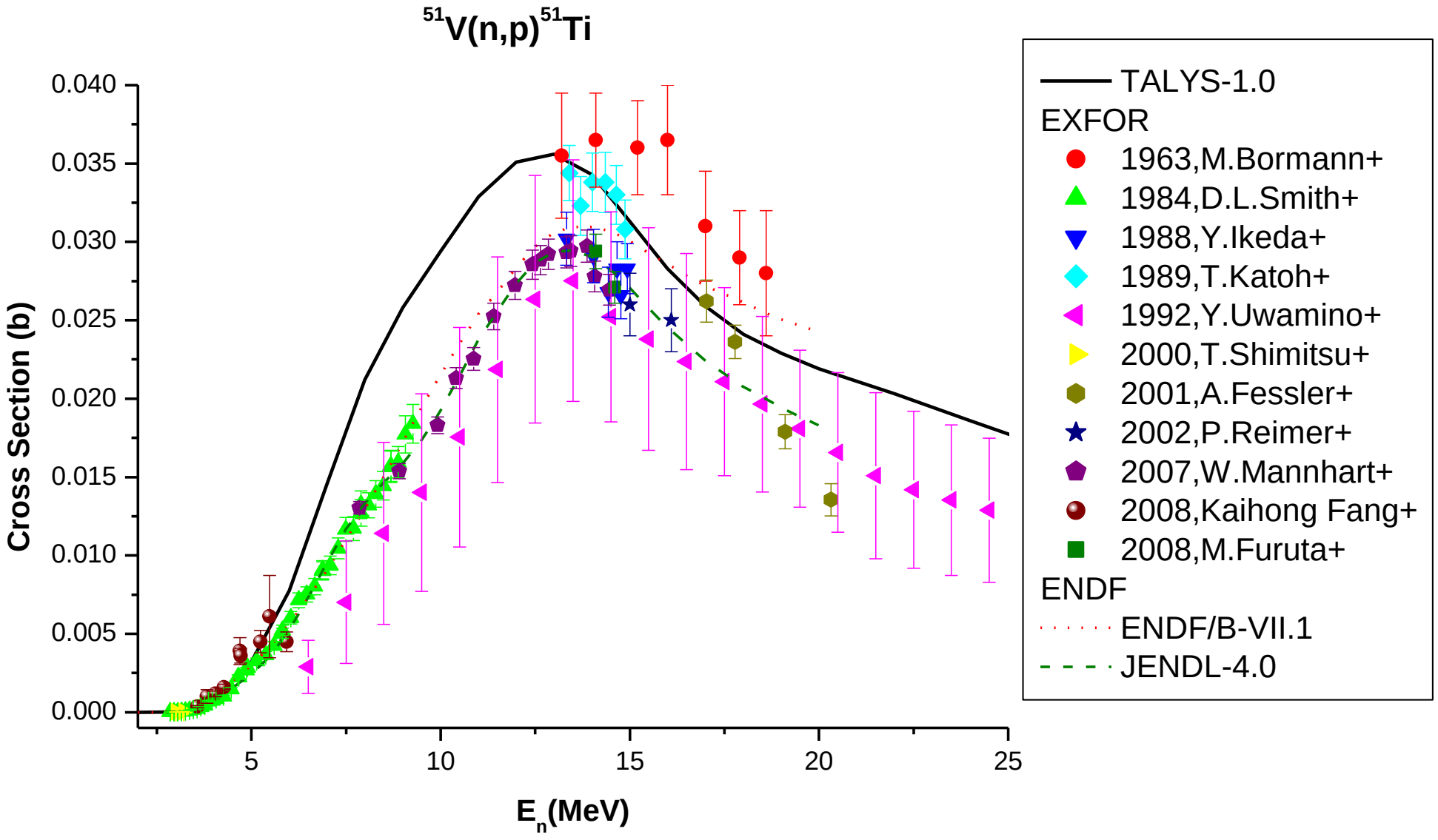

Fig. 2 The comparison of calculated excitation function using TALS-1.2 of $^{51}$V(n,p)$^{51}$Ti reaction with available experimental values and evaluated nuclear data files ENDF/B-VI, JENDL-3.3 and JEFF3.1.The values reported in Ref. (30)

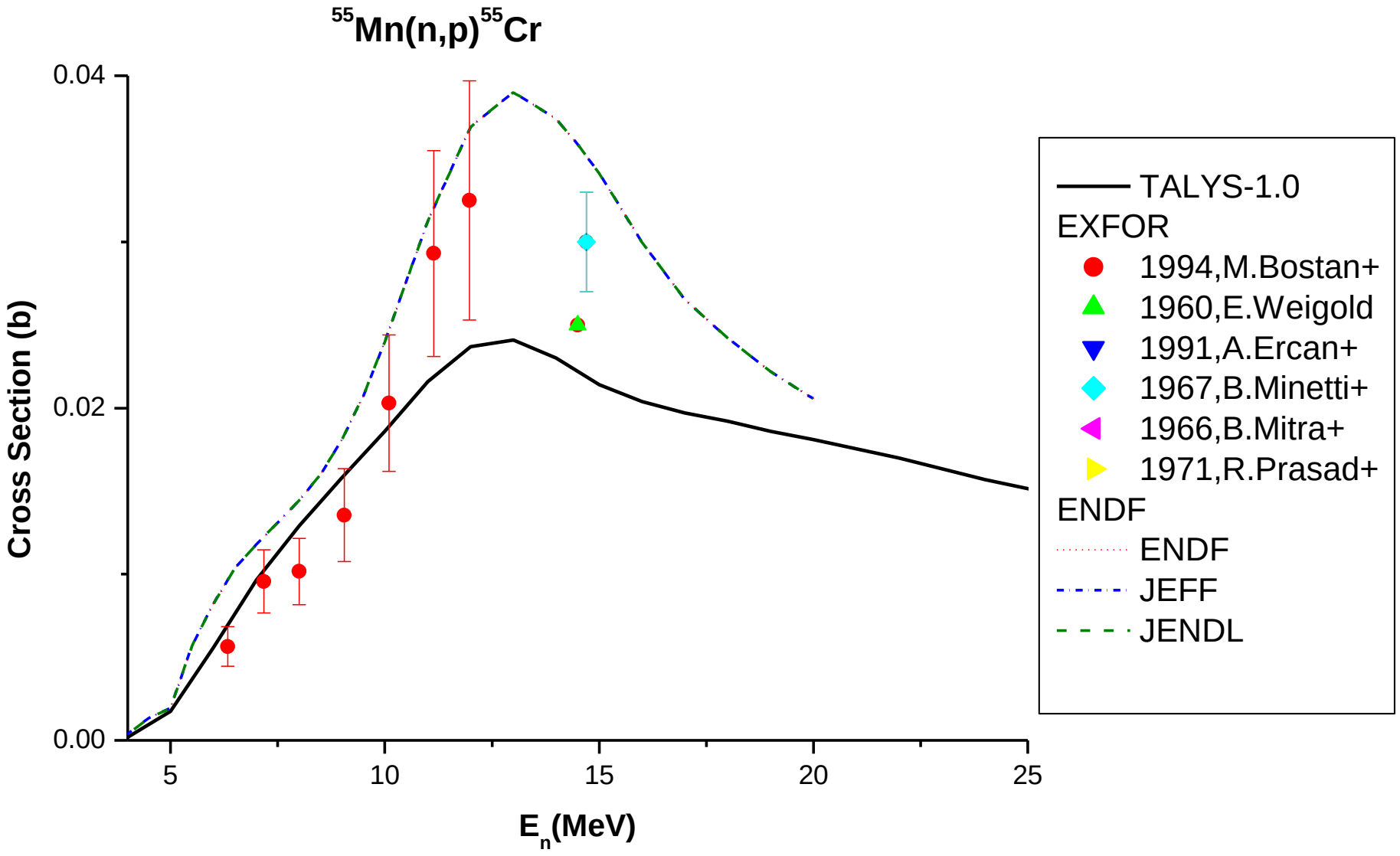

Fig. 3 The comparison of calculated excitation function using TALS-1.2 of $^{55}$Mn(n,p)$^{55}$Cr reaction with available experimental values and evaluated nuclear data files ENDF/B-VI, JENDL-3.3 and JEFF3.1.The values reported in Ref. (30)

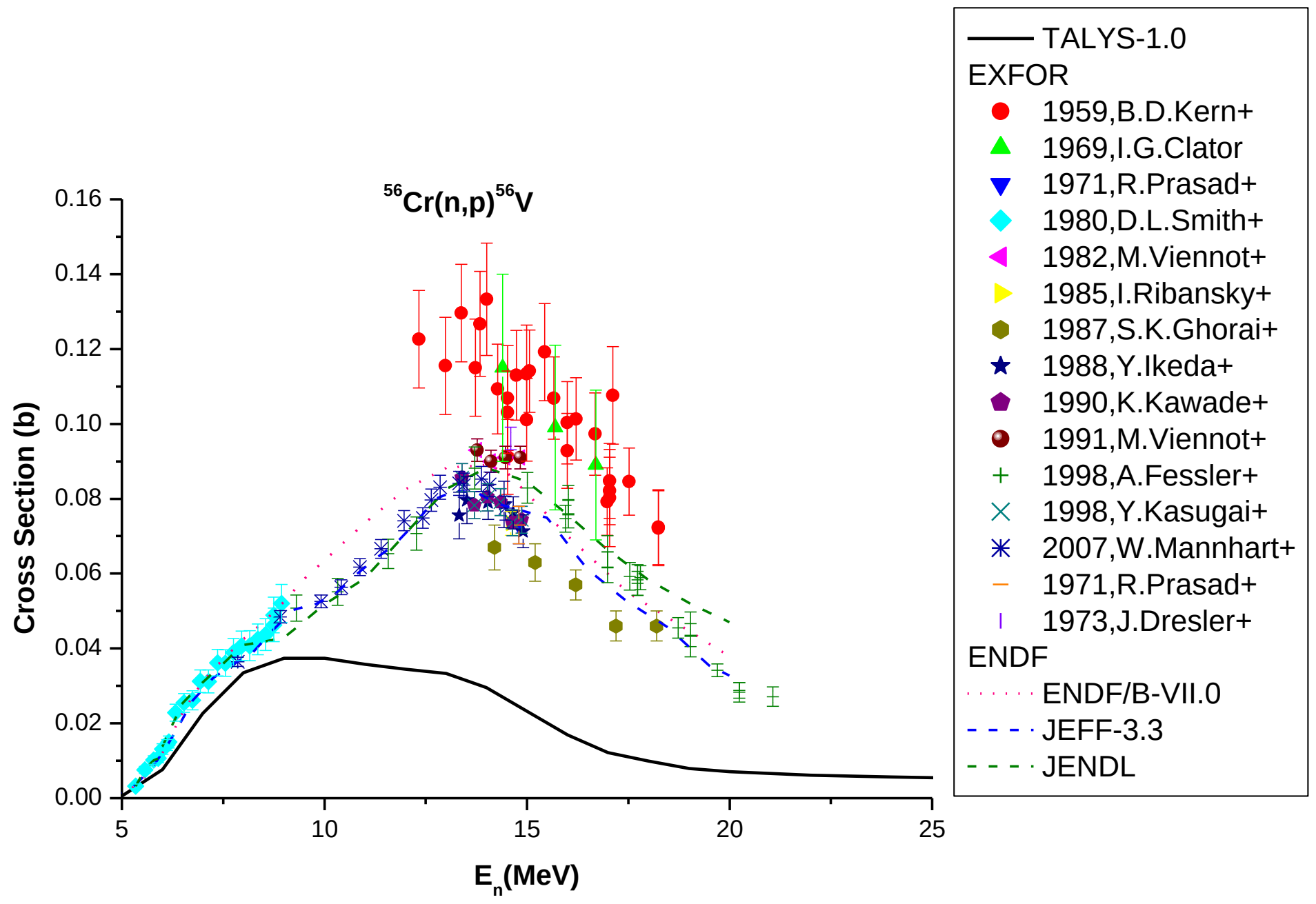

Fig. 4 The comparison of calculated excitation function using TALS-1.2 of $^{52}Cr(n,p)^{52}V$ reaction with available experimental values and evaluated nuclear data files ENDF/B-VI, JENDL-3.3 and JEFF3.1.The values reported in Ref. (30)

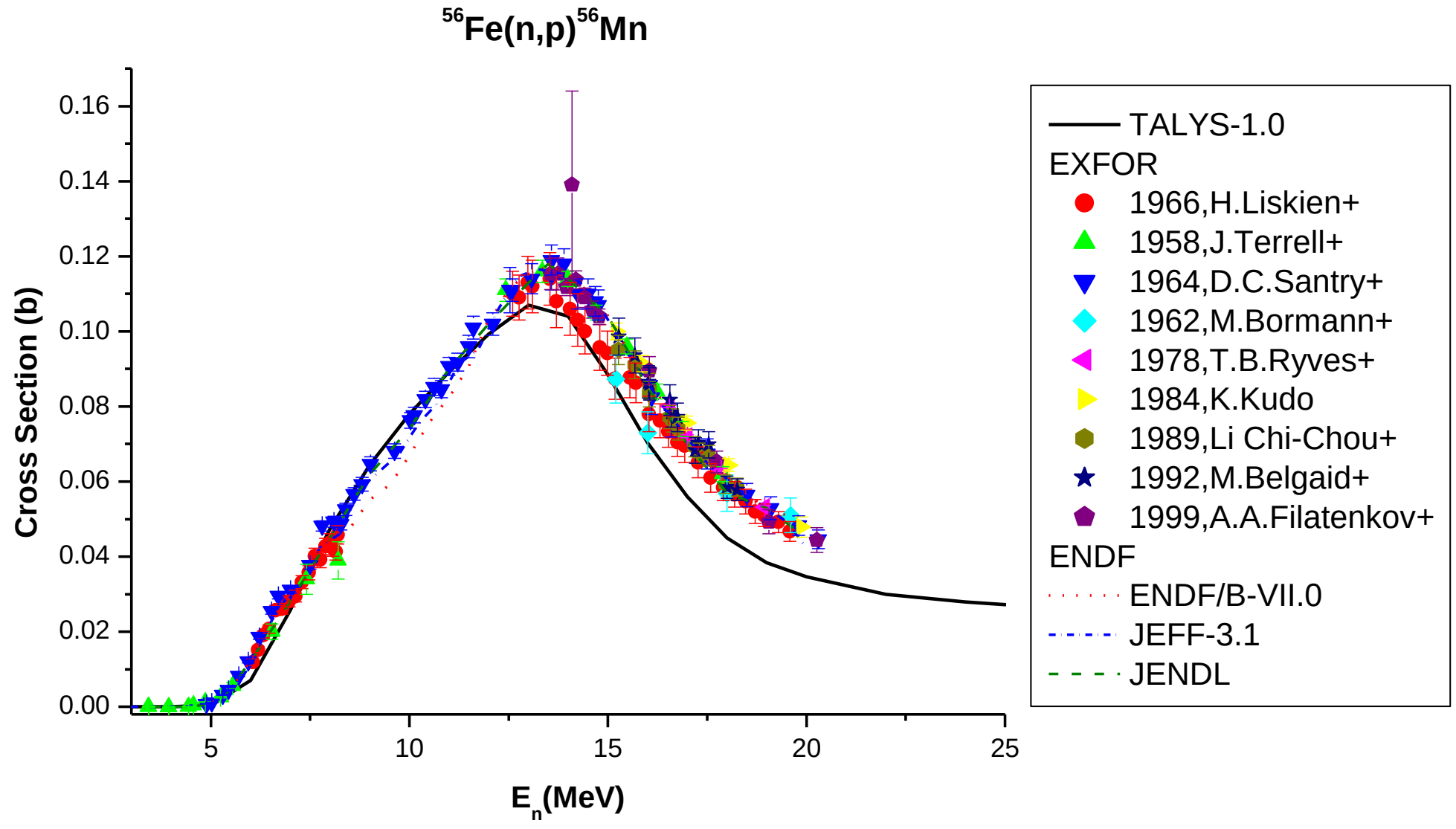

Fig. 5The comparison of calculated excitation function using TALS-1.2 of $^{56}Fe(n,p)^{56}Mn$ reaction with available experimental values and evaluated nuclear data files ENDF/B-VI, JENDL-3.3 and JEFF3.1.The values reported in Ref. (30)

## 3.Results and Discussions:

In present paper, (n, p) reactions for some structural fusion materials such as $^{27}$Al, $^{51}$V, $^{52}$Cr, $^{55}$Mn and $^{56}$Fe have been investigated in Figs. 1, 2, 3, 4 & 5. The new calculations on the excitation functions of *$^{27}$Al(n,p) $^{27}$Mg, $^{51}$V(n,p) $^{51}$Ti, $^{52}$Cr(n,p) $^{52}$V, $^{55}$Mn(n,p) $^{55}$Cr* and *$^{56}$Fe(n,p) $^{56}$Mn* reactions have been carried out up to 30 MeV incident neutron energy. The calculated results have been also compared with available experimental data in the literature and with ENDF/B-VII, T=300k; JENDL-3.3, T=300k and JEFF3.1, T=300k evaluated libraries . A reasonable agreement with experimental and theoretical excitation functions was obtained. The results can be summarized and conclude as follows:

### 3.1. $^{27}$Al(n,p)$^{27}$Mg Reaction

The experimental points are between 2 - 21 MeV, there is no experimental value between 21-25 MeV . ENDF/B-VI, JENDL-3.3 and JEFF3.1 files are in agreement with each other. There is excellent agreement between the cross-section calculated with TALYS-1.0 and the experimental data.

### 3.2. $^{51}$V(n,p)$^{51}$Ti Reaction

The experimental points are between 1 - 25 MeV, ENDF/B-VI and JENDL-3.3 files are in agreement with each other. There is good agreement between the cross-section between13-25 MeV calculated with TALYS-1.0 and the experimental data.

### 3.3. $^{55}$Mn(n,p)$^{55}$Cr Reaction

There are only alimited number of experimental cross-section data for $^{55}$Mn(n,p)$^{55}$Cr in the energy range up to 25 MeV. ENDF/B-VI, JENDL-3.3 and JEFF3.1 files are in agreement with each other. There is good agreement between the cross-section between 6-12 MeV calculated with TALYS-1.0 and the experimental data. The calculations between 12-25 MeV are not in good agreement with experimental data for $^{55}$Mn(n,p)$^{55}$Cr reaction .

### 3.4. $^{52}$Cr(n,p)$^{52}$V Reaction

The experimental points are between 1 - 20 MeV, ENDF/B-VI, JENDL-3.3 and JEFF3.1 files are in agreement with each other. It could not be said that the calculation are in good agreement with experimental data for $^{52}$Cr(n,p)$^{52}$V reaction.

### 3.5. $^{56}$Fe(n,p)$^{56}$Mn

The experimental points are in the energy range up to 20 MeV, ENDF/B-VI, JENDL-3.3 and JEFF3.1 files are in agreement with each other. There is good agreement between the cross-section calculated with TALYS-1.0 and the experimental data.